\documentclass[12pt]{iopart}

\usepackage{iopams}  
\usepackage{graphicx}
\usepackage{dcolumn}
\usepackage{bm}
\usepackage{color}
\usepackage{pbox}
\begin{document}

\title[Slip-velocity of large neutrally-buoyant particles in turbulent flows]{Slip-velocity of large neutrally-buoyant particles in turbulent flows}

\author{G Bellani and E A Variano}

\address{Department of Civil and Environmental Engineering, University of California, Berkeley, CA 94720, USA}
\ead{bellani@berkeley.edu}
\begin{abstract}
We discuss possible definitions for a stochastic slip velocity that describes the relative motion between large particles and a turbulent flow.  This definition is necessary because the slip velocity used in the standard drag model fails when particle size falls within the inertial subrange of ambient turbulence.  We propose two definitions, selected in part due to their simplicity: they do not require filtration of the fluid phase velocity field, nor do they require the construction of conditional averages on particle locations.  A key benefit of this simplicity is that the stochastic slip velocity proposed here can be calculated equally well for laboratory, field, and numerical experiments.  The stochastic slip velocity allows the definition of a Reynolds number that should indicate whether large particles in turbulent flow behave (a) as passive tracers; (b) as a linear filter of the velocity field; or (c) as a nonlinear filter to the velocity field.  We calculate the value of stochastic slip for ellipsoidal and spherical particles (the size of the Taylor microscale) measured in laboratory homogeneous isotropic turbulence.  The resulting Reynolds number is significantly higher than 1 for both particle shapes, and velocity statistics show that particle motion is a complex non-linear function of the fluid velocity.  We further investigate the nonlinear relationship by comparing the probability distribution of fluctuating velocities for particle and fluid phases.  

\end{abstract}

\maketitle

\section{Introduction}\label{sec:1}
It is well known that particles suspended in a turbulent flow can significantly modulate the turbulent dynamics even at moderate concentrations \cite{Balachandar:2010hp,Toschi:2009fi,Poelma:2006bc}.
The ability to predict turbulence modulation by particles (\emph{e.g.} changes in turbulent kinetic energy or its dissipation rate) as well as the dynamics of the suspended particles (\emph{e.g.} aggregation or dispersion) strongly depends on our ability to predict the forces that the individual particles experience in turbulent flows \cite{Squires:2011up,Elghobashi:1993vi}. 
 
However, quantifying the forces that mediate fluid-particle coupling is nontrivial.  Neither the stochastic nor deterministic description of turbulent suspensions is currently at the level required for useful engineering predictions.  The limits of current knowledge were summarized eloquently by Qureshi \emph{et al.}  \cite{Qureshi:2007gx} : ``[...] writing (not to mention solving) the particle equation of motion in the most general case remains a challenge, and only limited cases are treated at present.''

The forces which explicitly couple fluid and particle phases are the shear and normal stresses felt at the particle-fluid boundary.  Many prediction schemes attempt to simplify the description by replacing the integral of surface forces with a local body force,  the drag force $F_D$. 
Analytical expressions for this force can be derived only in very restrictive conditions (\emph{e.g.} zero Reynolds number, slow relative motion, simple shear, etc.). The range of application of these solutions can sometimes be extended by adding empirical coefficients (\emph{e.g.} Oseen correction Shiller and Neuman formula, for drag coefficient of spheres at finite Reynolds number \cite{Clift:2005vm}).

Such extensions work well when they describe cases that do not deviate too greatly from the conditions for which the original models are derived, so that the empirical coefficients can compensate for the errors.  However, empirical coefficients can no longer help when the physics of the system to be modeled differs greatly from the analytical; a harbinger of this failure is when concepts or ideas that are central to the model start to be ill--defined.  Such is the case for large particles in turbulent flow: with increasing size, the slip velocity becomes ill--defined \cite{Lucci:2010gc}, and no simple extensions of the standard drag model have been successful at predicting the drag experienced by particles (reviewed further in section \ref{sec:2}).  Our goal in this paper is to specify a stochastic slip velocity that remains well--defined for arbitrarily-shaped large particles in turbulent flows, and can be computed in a straightforward manner.  Such a definition will be useful in the community's attempts to formulate stochastic drag models predicting particle--fluid coupling, and also in defining non-dimensional numbers that parametrize other aspects of the flow (e.g. particle clustering or collision rates).


\section{Background}\label{sec:2}
Slip velocity $\mathbf{U_s}$ is an essential parameter in models for the behavior of turbulent suspensions.  Slip velocity commonly appears as the velocity scale in the particle Reynolds number: $Re_p\equiv V d_p\nu^{-1}$.  It is also used in drag models that predict the coupling of particle and fluid phase motion.  To understand some definitions of slip velocity, and opportunities to improve it, we review the drag models used for single particles in flows of increasing complexity.

\subsection{Drag models}
The common origin of most of the drag models used to describe particle motion in unsteady flow comes from the exact solution of the unsteady Stokes equation for a particle that oscillates relative to a homogeneous quiescent flow (or is held fixed in an oscillating flow) \cite{Clift:2005vm}. 
\begin{equation}
-\mathbf{F_D}=3\pi\mu d_p\mathbf{U}+\frac{3}{2}(\pi\rho_f\mu)^{1/2}\int_{-\infty}^t\left(\frac{d\mathbf{U}}{dt}\right)(t-s)^{1/2}ds+\frac{2}{3}\pi d_p^3 \rho_f\frac{d\mathbf{U}}{dt},
\label{eq:fd_o}
\end{equation}
where $d_p$ is particle diameter and $\rho_f$ is the fluid density. $\mathbf{U(t)}$ is the relative velocity between particle and fluid, which is trivial to determine given that one phase is defined as steady while the other has its motion prescribed exactly.  Eq.~(\ref{eq:fd_o}) contains the following three terms on the right hand side, in order: a steady term (Stokes drag), a history term (augmented drag), and an added mass. 
Many attempts have been made to generalize eq.~(\ref{eq:fd_o}) to account for more complicated motions, as discussed in \cite{Clift:2005vm}. One of the most successful is the expression derived by Maxey and Riley \cite{Maxey:1983wj}, which is often applied in the form:
\begin{equation}
-\mathbf{F_D}=3\pi\mu d_p\mathbf{U_s}+\frac{3}{2}(\pi\rho_f\mu)^{1/2}\int_{-\infty}^t\left(\frac{d\mathbf{U_s}}{dt}\right)(t-s)^{1/2}ds+\frac{2}{3}\pi d_p^3 \rho_f\frac{d\mathbf{u_f}}{dt},
\label{eq:MR}
\end{equation}
where the Fax\'en terms that account for (linear) velocity gradients across particle scale are neglected.  
All terms now depend on the slip velocity $\mathbf{U_s}\equiv \mathbf{u_p}(\mathbf{x}_{cg},t)-\mathbf{u_f}(\mathbf{x}_{cg},t)$, where $\mathbf{x}_{cg}(t)$ is the location of the particle center and $\mathbf{u_f}(\mathbf{x}_{cg}, t)$ is the ``undisturbed'' fluid--phase velocity corresponding to location of the particle center.  This definition of slip velocity is only valid in the limit of small particle size and particle Reynolds number, for this makes it possible to interpolate a fluid velocity at the particle center.  These restrictions on size and speed also underlie the derivations of eq.~(\ref{eq:fd_o}) and eq.~(\ref{eq:MR}).  

Although a turbulent flow is far more complicated than a simple oscillatory motion, eq.~(\ref{eq:MR}) successfully describes the motion of spherical particles suspended in turbulence under certain conditions \cite{Anonymous:D-LJEn02,Balachandar:2010hp}.  Specifically, particles must be small enough and changes in relative motion slow enough that key assumptions in eq.~(\ref{eq:MR}) are met;   velocity gradients must be at most linear at their scale ($d_p$); in turbulent flow this condition can be expressed as $d_p/\eta \ll 1$, where $\eta$ is the Kolmogorov scale.  The particles must also move slowly enough relative to the surrounding fluid that $Re_p<1$.  When $Re_p>1$ the Stokesian drag in eq.~(\ref{eq:MR})  becomes inaccurate.  For the special case of small spherical particles at high speeds, the nonlinearity can be corrected with empirical factors such as $\beta=(1+0.15Re_p^{0.687})$ \cite{Clift:2005vm}, which
appear in steady part of the drag model as $F_S=3\pi\mu d_p \beta U_s$.
For large particles, it is not as easy to account for nonlinearities.  

When the flow is significantly nonuniform over the particle surface, \emph{e.g.} for particles larger than the Kolmogorov lengthscale, there is not currently an equation that successfully predicts the forces coupling the fluid and a freely moving particle. Both eq.~(\ref{eq:fd_o}) and (\ref{eq:MR}) depict particle motion that applies a temporal filter to the ambient flow.    Neither, however, include an equivalent spatial filter.  Such a spacial filter seems like a necessary step for describing the forces on large particles \cite{Calzavarini:2009ub}.  One way to include this filtering effect in eq.~(\ref{eq:MR}) is to apply the Fax\`en corrections in the form volume and surface averages applied to the fluid phase at the scale of the particle. This approach is able to qualitatively reproduce some aspects of acceleration statistics  \cite{Calzavarini:2009ub,Volk:2011du}, but does not provide a satisfactory method for quantitative prediction of particle motion.   

Such an extension may simply require more tuning to be successful, or it may be that the simple physics used in deriving in eq.~(\ref{eq:fd_o}) and (\ref{eq:MR}) cannot be directly extended to the most general case.  A key issue in guiding this model to success, or understanding its failure, is defining a meaningful slip velocity.  That is,  the drag terms in eq.~(\ref{eq:fd_o}) and (\ref{eq:MR}) use a single velocity scale to represent the relative motion of the particle and fluid, while it is unclear which single velocity (if any) can parameterize the drag on large particles.  
\subsection{Slip velocity: current extensions}
In section \ref{sec:sv} we propose a new definition of slip velocity specifically designed for the case of large particles in turbulent flow.  Before discussing this, we review three possible definitions of slip velocity that have been used when attempting to extend eq.~(\ref{eq:MR}) to large particles.  Each definition uses a different fluid velocity, and they all share the same choice of particle velocity, namely the instantaneous center of mass velocity.  The three choices are: 
$U_{s(1)}$ interpolate the fluid velocity to the particle center, 
$U_{s(2)}$ average the fluid velocity at the particle wall, and 
$U_{s(3)}$ average the fluid velocity at scales larger than the particle.  

$U_{s(1)}$ is the definition used in eq.~(\ref{eq:MR}), and is straightforward to compute for small particles, but it is questionable whether this approach is physically meaningful when velocity gradients change significantly on the scale of the particle, and/or particles have a complex shape.  Furthermore, when $d_p>\eta$,  it becomes impossible to exactly interpolate the fluid field to the particle center.  Two different numerical methods offer a way work around this impossible task and use $U_{s(1)}$.  First, if a particle does not disrupt the fluid phase, but only follows it via one-way coupling, then the fluid phase velocity can be determined anywhere \cite{Calzavarini:2012fq}.  Second, a particle-free realization of the flow with identical initial and boundary conditions can be used to assign fluid velocities corresponding to the location of particle centers in a particle-laden simulation \cite{Bagchi:2003ey}. Obviously, neither of these methods are practical in laboratory studies.

$U_{s(2)}$ is possible only in certain idealized numerical approaches, because in the real world, $U_{s(2)}$ is strictly zero because of the no--slip condition.  The two special numerical cases discussed above that make it possible to use $U_{s(1)}$ also allow $U_{s(2)}$ to be used.  $U_{s(2)}$ determined from a one--way coupled simulation is the definition used in the work of \cite{Calzavarini:2012fq}.

$U_{s(3)}$ can be used in both numerical and laboratory experiments, as it is based on the fluid flow outside of the particle volume.  The complication with this method is that one must specify a lengthscale over which to average the fluid flow.  This filter lengthscale then becomes a new parameter in the model, and includes an inherent choice about which scales of the flow are important to particle motion.  This choice can be complicated in case of anisotropic particles, since multiple particle lengthscales are involved. There is also a practical challenge to using this definition to analyze laboratory data, for the averaging requires three--component velocity measurements over a three-dimensional measurement volume.

Few studies have evaluated the performance of eq.~(\ref{eq:MR}) in predicting the motion of large particles in turbulence, using all three slip velocity definitions discussed above.  Bagchi \& Balachandar   \cite{Bagchi:2003ey} use $U_{s(1)}$ and $U_{s(3)}$ to predict the drag on a particle that is fixed relative to the surrounding flow, and use DNS to compute the flow and the exact drag force on the particle.  They find that eq.~(\ref{eq:MR}) does not predict the instantaneous drag force accurately for large particles.  The disagreement between measurement and model was not remedied by changes in the choice of slip velocity, nor by the inclusion of Fax\'en correction, and it increases with increasing turbulence level. They also find that the mean value is best captured by the steady term only. 

Calzavarini \emph{et al.}  \cite{Calzavarini:2012fq} use $U_{s(2)}$ to predict the drag on a neutrally buoyant particle moving freely in a turbulent flow, a case that is more general than that of \cite{Bagchi:2003ey}. They assess the performance of eq.~(\ref{eq:MR}) (with Fax\'en forces) by comparing particle velocity and acceleration statistics with those from DNS and laboratory data.  
They find that eq.~(\ref{eq:MR})  can reproduce some of the statistics of particle kinematics, but not others.  By removing different terms from the model, they find that inertial terms with Fax\'en correction are essential to describe acceleration statistics, whereas the qualitative behavior of velocity statistics are mainly determined by the non-linear steady drag term. 
The fact that eq.~(\ref{eq:MR}) fails to reproduce velocity fluctuations of large particles has been confirmed also in laboratory experiments \cite{Sapsis:2011ii}. 

In conclusion, both studies agree that eq.~(\ref{eq:MR}), in its current state, is insufficient for predicting the forces on large particles in turbulent flow.  The results do not clearly indicate which terms are failing, but they do emphasize the importance of the steady drag term.  As a result, we conclude that any opportunity to improve this drag term is worth pursuing.  We focus on the slip velocity as a parameter that must be well-defined and physically meaningful before any drag model can be successful.  As the three definitions discussed above each have significant limitations, we propose a new definition of slip velocity herein.  Unlike $U_{s(1)}$ and $U_{s(2)}$, our can be applied in both numerical and laboratory studies, and unlike $U_{s(3)}$ it does not require a specific lengthscale.  It can be used to formulate stochastic drag models (as suggested in \cite{Balachandar:2010hp,Sapsis:2011ii}), to compare the results of laboratory measurements and numerical models, and to compute particle Reynolds numbers.  It is this last application that motivated us to consider this definition.  In our study of arbitrarily-shaped large neutrally--buoyant particles in turbulence with zero mean flow, we wanted a straightforward definition of particle Reynolds number that could help us quantify the transition from passive tracers to more complex behavior.  Herein, we demonstrate the our definition of slip velocity achieves this goal.



\section{Stochastic slip velocity}\label{sec:sv}

We begin with two possible choices for the stochastic slip velocity $\mathbf{V}$:

\begin{equation}\label{eq:usl1}
{\mathbf{V_A}} \equiv {\left( {\langle \mathbf{u_f'}^2\rangle  - \langle \mathbf{u_p'}^2\rangle } \right)^{1/2}}
\label{eq:VA}
\end{equation}
and
\begin{equation}\label{eq:usl2}
{\mathbf{V_B}} \equiv {\langle {(\mathbf{u_f'}  - \mathbf{u_p'} )^2}\rangle ^{1/2}}.
\label{eq:VB}
\end{equation}
Here, angle brackets represent the expectation value (obtained by appropriate averaging) and primes represent the fluctuation relative to the expectation value ($\mathbf{u'}\equiv\mathbf{u}-\langle\mathbf{u}\rangle$).
The definitions for $\mathbf{V_A}$ and $\mathbf{V_B}$ intentionally remove any constant difference in the mean velocity of fluid and particle phases.  The slip due to such differences can be characterized easily and considered the steady counterpart to this stochastic slip velocity ($\mathbf{V_s}\equiv\langle\mathbf{u_f}\rangle-\langle\mathbf{u_p}\rangle$).
While all of the slip velocities discussed above are vector quantities, drag models are typically applied component--wise, and thus we can simplify the forthcoming discussion by considering only a single component.  Thus we reduce vector quantities (boldface) to a single velocity component (regular typeface).  

The definitions $V_A$ and $V_B$ have slighlty different practical advantages.  One immediate advantage of $V_A$ is that it  can be determined by measuring the fluid and particle phase velocity statistics independently.  In contrast, $V_B$ includes a covariance $\langle{u_f}\prime{u_p}\prime\rangle$.     
To evaluate this covariance requires the work-around strategies discussed above.  
Because $V_A$ does not require such work-around, then it can be evaluated  directly without introducing any spatial filtering and/or arbitrary length scale.  
A benefit of $V_B$ is that it arises naturally from eq.~(\ref{eq:MR}) if a Reynolds' decomposition is performed on $U_s$, therefore it seems a natural candidate for the definition of a stochastic slip. 

To understand the relationship of $V_A$ and $V_B$, we consider two extreme cases.  A passive tracer, by definition, will have $\langle u_p'^2\rangle=\langle u_f'^2\rangle=\langle u_f'u_p' \rangle$.  In this case, both stochastic fluctuating slip velocities will be zero: $V_A=V_B=0$.  The other extreme is a particle that does not respond to any of the fluctuating fluid velocities.  This may be because the particle is anchored at a location, or because it moves ``ballistically'' through the flow with large inertia.  By this definition, this extreme case has $u_p'=0$, which makes $\langle u_f'u_p' \rangle=0$ as well.  Thus in this extreme case both our definitions of slip velocity equal the magnitude of velocity fluctuations in the fluid phase: $V_A=V_B=\langle u_f'^2\rangle^{1/2}$.  Both of these limiting behaviors (totally passive and totally ballistic) yield stochastic fluctuating slip velocities that make good sense: passive particles do not slip relative to the fluid phase, while 100\% of the turbulent fluctuations slip past ballistic particles.  

We can also use these limiting behaviors to provide a constraint that can help us prescribe models for $\langle u_f'u_p'\rangle$.  We start with the assumption that $\langle u_f'u_p' \rangle=f(\langle{u_f'}^2\rangle,\langle{u_p'}^2\rangle)$.  Nondimensionalizing by $\langle{u_f'}^2\rangle$ gives $\frac{\langle{u_f}'{u_p}' \rangle}{\langle{u_f'}^2\rangle}=F(\frac{\langle{u_p'}^2\rangle}{\langle{u_f'}^2\rangle})$.  The limiting behaviors discussed above imply that $F(0)=0$ and $F(1)=1$.  The simplest function satisfying these constraints is linear: $F(x)=x$.  This choice implies that the covariance equals the particle velocity variance, which makes the definitions for $V_A$ and $V_B$ identical.  Until proper data is available to support a more complex model for the covariance, we will assume that this simple model is acceptable.  This leaves us with a single definition of stochastic slip velocity: $V=V_A\approx V_B$, which can be easily evaluated in laboratory, field, and numerical experiments according to eq.~(\ref{eq:VA}).  

Note that $V\equiv V_A$ can be easily visualized as the difference between the power spectrum of particle velocity and the power spectrum of the fluid phase velocity. Hence, from the physical point of view, $V_A\approx V_B$ can be seen as a consequence that large particle dynamics tend to be dominated by fluid motion at scales much larger than particle size \cite{Kuboi:1974vg}, thus the difference between particle and fluid variance is a reasonable choice for a stochastic measure of the slip velocity. 
We analyze $V$ using laboratory data below.

\section{Experiment}\label{sec:exp}
\subsection{Facility}
\begin{figure}
\centering
\includegraphics[trim=0cm 0cm 0cm 0.85cm, clip=true, width=0.85\textwidth]{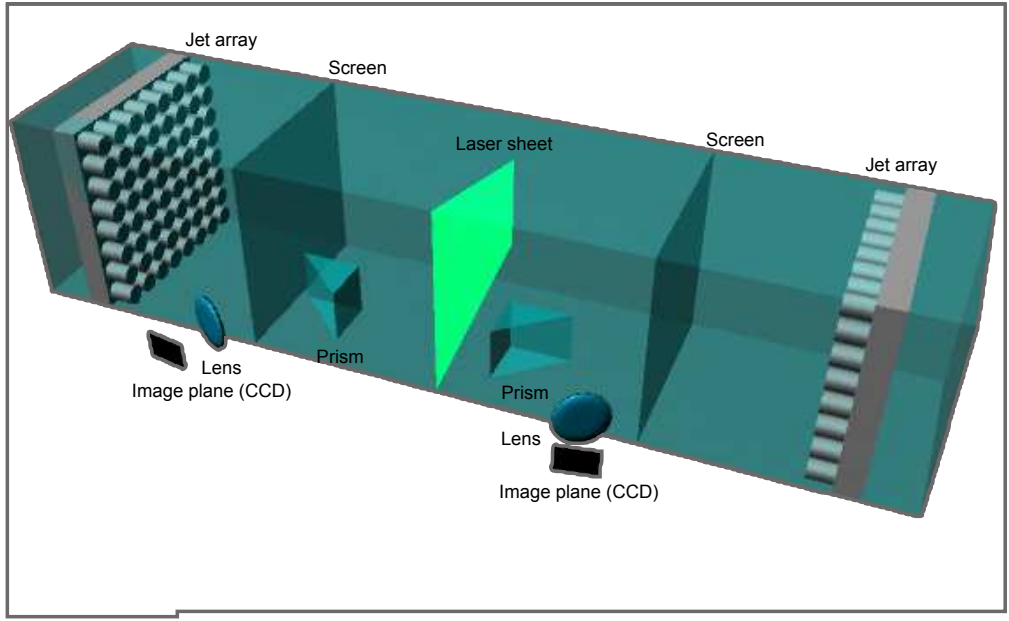}\\
\caption{Basic design of stirred turbulence tank and experimental setup for particle imaging.  The 64 pumps on each wall fire in a stochastic pattern to maximize Reynolds number and minimize mean flow.}\label{fig:tank_setup}  
\end{figure}

Experiments are performed in a turbulent water tank, described in detail in \cite{Bellani:2012ty,Bellani:2012wn,Variano:2008dn}. The flow is driven by two symmetric arrays of randomly--firing synthetic jets, which create homogeneous isotropic turbulence with almost zero mean flow. The idea is illustrated in figure \ref{fig:tank_setup}. 
The resulting flow is high--Reynolds--number turbulence that is both homogeneous and isotropic over a large volume at the center of the tank.  Importantly, this central region of idealized turbulence is significantly larger than the integral lenghtscale of turbulence, from which we conclude that the particle kinematics measured at the tank center are caused by a single set of turbulent forcing parameters.  That is, they do not carry a strong signature of flow from other regions via the history term in eq.~(\ref{eq:MR}).

\begin{figure}[tbp]
\centering
\includegraphics[width=0.75\textwidth]{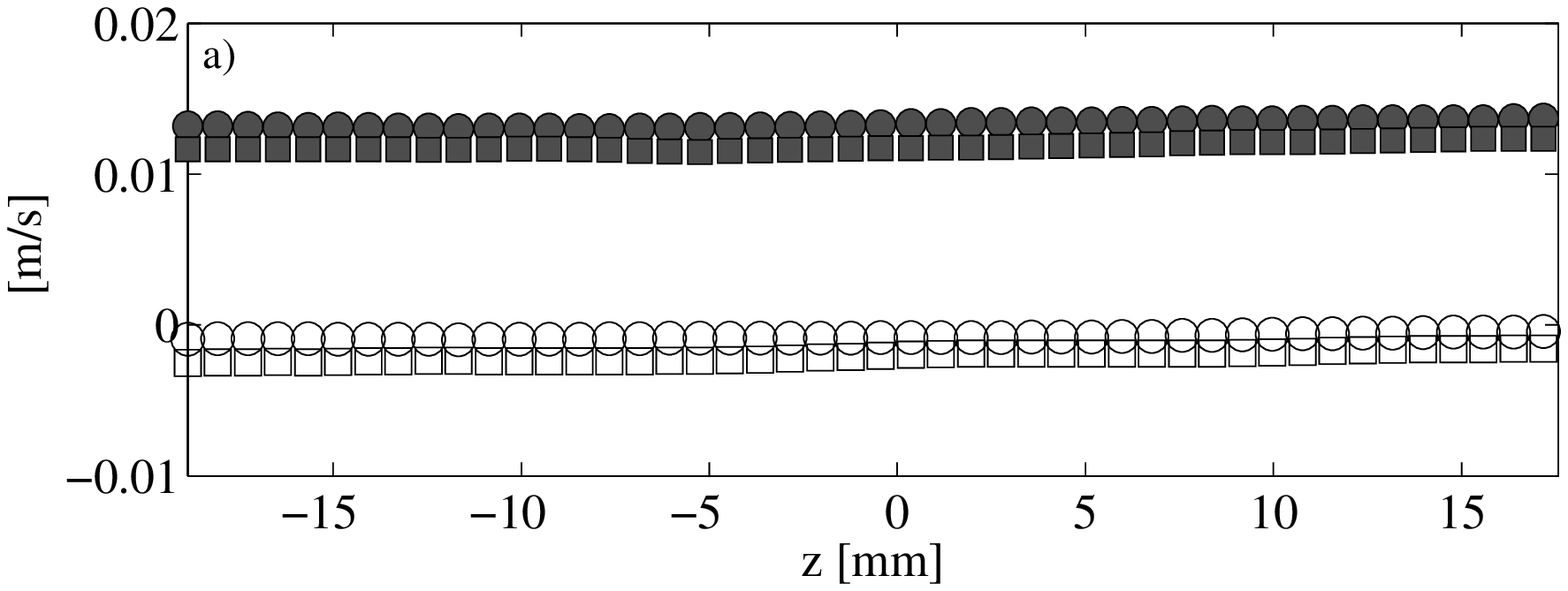}\\
\includegraphics[width=0.75\textwidth]{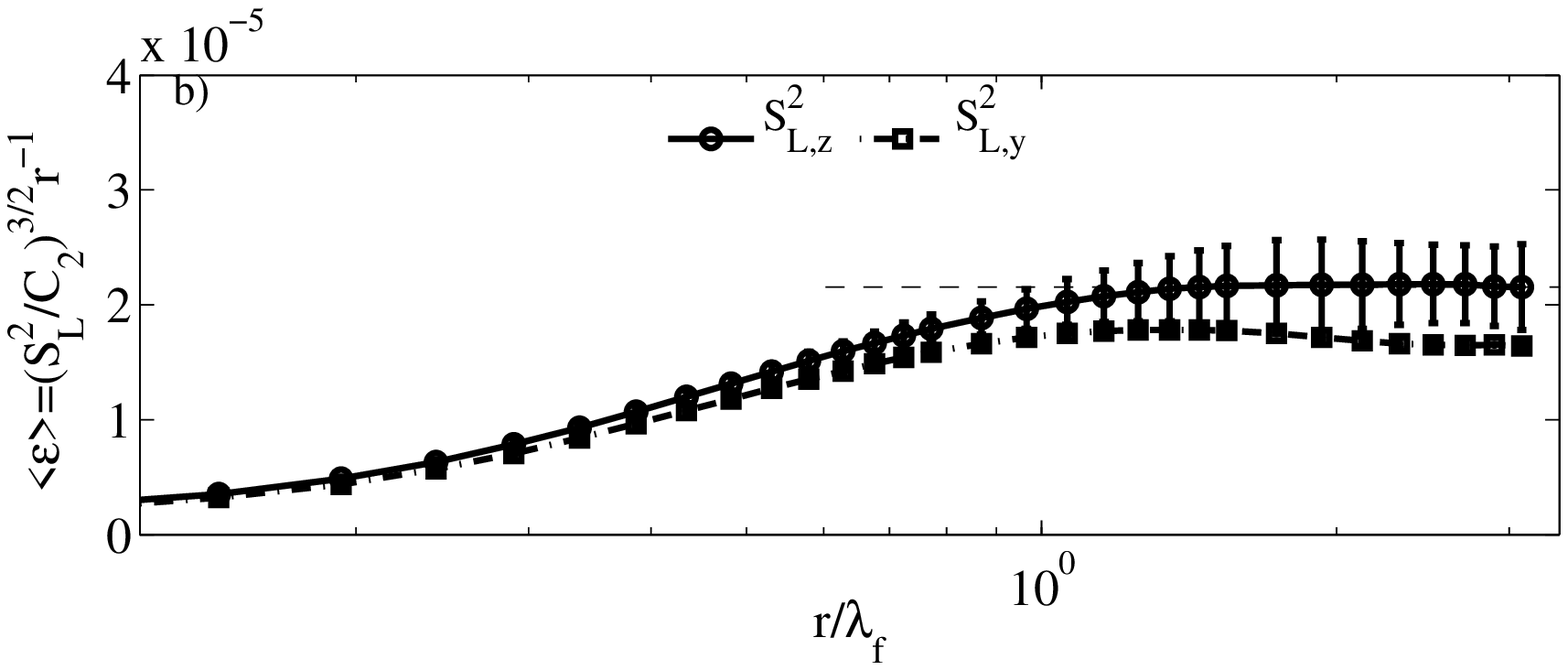}
\caption{[a) velocity variance (gray) and mean velocity (white) measured across space (only every third sample in space is shown).  Velocity components in $y$ (square) and $z$ (circle) show near isotropy and complete homogeneity over the spatial region shown.  Marker sizes are representative of the 95\% confidence intervals. b) second order longitudinal velocity structure functions for $y$ (square) and $z$ (circle), shown in compensated form.  95\% confidence intervals are shown for $y$ only, and are very similar for $z$.}\label{fig:isotropy}  
\end{figure}

For the experiments reported herein, we have modified the tank relative to that which is described in \cite{Bellani:2012ty,Bellani:2012wn,Variano:2008dn}.
We have added two screens with 5mm mesh in the $x-y$ plane, to help confine the particle motion to the central region of the tank.  This change was primarily for experimental convenience.  The velocity statistics in this new configuration are reported in figure \ref{fig:isotropy} and table \ref{table:turb}.  They are measured by 2D2C PIV in the $y-z$ plane.  We choose this plane because the symmetry of the tank should make $x$ and $y$ similar, while the flow in the axial ($z$) direction may be different.  Figure \ref{fig:isotropy} shows that velocity statistics in the $z$-direction are actually quite similar to those in $y$.  Specifically, figure \ref{fig:isotropy}a shows that the velocity variance (a large-scale quantity) is nearly isotropic, and homogeneous over a region of at least 4cm in $z$.  Figure \ref{fig:isotropy}a also shows that the mean velocity is much smaller than the typical fluctuating velocities.  Figure \ref{fig:isotropy}b shows the second--order longitudinal velocity structure functions computed in the $y$ and $z$ directions.  The difference between these curves is proportional to the anisotropy in velocity statistics, and by examining the difference over $r$, one can make a scale--by--scale comparison of isotropy.  As expected from the Kolmogorov hypotheses, the anisotropy is maximum at large scales (large $r$) and decreases until the flow is completely isotropic by the smallest measured scales.  Another use of the structure function is to measure the dissipation rate of turbulent kinetic energy ($\varepsilon$).  Kolmogorov predicts that $S_L^2=C_2\varepsilon^{2/3} r^{-2/3}$ for $r$ values within the inertial subrange of turbulence.  This is exactly what is seen in Figure \ref{fig:isotropy}b, for which we use the common value of $C_2=2$.  To determine the bounds of the inertial subrange, we calculate the velocity autocovariance across space (not shown), which gives the Taylor lengthscale $\lambda_f$ and the longitudinal integral lengthscale $\Lambda_f$.  These and related results are seen in Table \ref{table:turb}, with 95\% confidence intervals computed via bootstrap \cite{efron_introduction_1994}.

\begin{table}
\begin{center}
\begin{tabular}{rrcc}
\hline  
 &  &  & \underline{95\% CI} \\
Vertical velocity fluctuations, $\langle u_f\prime^2\rangle^{1/2}$ &[$\times 10^{-2}$ m s$^{-1}$] & $1.20$ & [1.14\ 1.26]\\
Axial velocity fluctuations,  $\langle u_{f(A)}\prime^2\rangle^{1/2}$ &[$\times 10^{-2}$ m s$^{-1}$] & $1.33$ & [1.19\ 1.42] \\
Transverse velocity fluctuations,  $\langle u_{f(T)}\prime^2\rangle^{1/2}$ &[$\times 10^{-2}$ m s$^{-1}$] &$\approx1.20$ & (\emph{by $x-y$ symmetry})\\
\\
Taylor microscale, 	$\lambda_f$     &[$\times 10^{-3}$ m] & 8.3 & [6.2\ 10]\\
Integral length-scale,  $\Lambda_f$ & [$\times 10^{-3}$ m] &  57 & [57\ 58]\\
Eddy turnover time,  $T$ ($=\Lambda_x\langle u_f\prime^2\rangle^{-1/2}$) & [s] & $4.3$ & [4.1\ 4.5]\\
\\
Kinematic viscosity,  $\nu$ &(m$^2$ s$^{-1}$) & $9.47\times10^{-7}$ & (at 22.6 Celsius) \\
Turbulent kinetic energy,  $k$  &[$\times 10^{-4}$ m$^2$ s$^{-2}$] & 2.28 & [1.91\ 2.57]\\
Turbulent dissipation rate,  $\varepsilon$  &[$\times 10^{-5}$ m$^2$ s$^{-3}$] & 2.15 & [1.82\ 2.56]\\
Kolmogorov length-scale,  $\eta$ ($=(\nu^3/\varepsilon)^{1/4}$) & [$\times 10^{-3}$ m] & 0.44 & [0.38\ 0.42]\\
Kolmogorov time-scale,  $\tau_{\eta}$ ($=(\nu/\varepsilon)^{1/2}$) & [s] & $0.16$ & [0.16\ 0.17]\\
\\
$\mbox{Re}_{\lambda}$ ($=\langle u_f\prime^2\rangle^{1/2}\lambda_x/\nu$)& & 115 & [81\ 140]\\
 $\mbox{Re}_{\Lambda}$ ($=\langle u_f\prime^2\rangle^{1/2}\Lambda_x/\nu$)& & 795\ & [750\ 840]\\
\hline
\end{tabular}
\end{center}
\caption{Turbulent statistics in measurement volume at the center of the stirred turbulence tank. Definitions are given in the table or in the text.}
\label{table:turb}
\end{table}


\subsection{Particles}
A suspension of large near--neutrally buoyant particles is studied at a volume fraction of 0.14\%.  Particles' density $\rho_p=1020 \textrm{kg/m}^3$ at 20$^\circ$C and they are manufactured by hand using injection molding of Agarose hydrogel solution.  These hydrogel particles are nearly transparent, and are closely matched to the refractive index of water.  Each particle contains hundreds of small ($\approx40$ $\mu$m) optical tracers, which we track to measure the velocity at a collection of points within each particle (see next section).
 
Two particle shapes are considered here: spheres of diameter $d_s=8$ mm, and prolate ellipsoids whit polar and equatorial axes $l_e = 16$ and $d_e= 8$ mm, respectively.  
These particle sizes are within the inertial subrange of ambient turbulence, between the Kolomgorov microscale 
and the integral lengthscale 
, and of similar size to the Taylor microscale 
. More details are given in table \ref{table:particles}.  

\begin{table} 
\centering
\begin{tabular}{lcccccc}
\hline
   & $(d_p,l_p)/\eta$ &  $(d_p,l_p)/\lambda_f$  &  $(d_p,l_p)/\Lambda_f$ &  $\tau_p/\tau_\eta$ &  $\tau_p/T$\\
\hline
 \emph{Spheres}    &  21, 21  & 0.65, 0.65     &  0.11, 0.11     &  26 & 1.1  \\
 \emph{Ellipsoids}  & 21, 42  & 0.65, 1.3   & 0.11, 0.22   &  39     & 1.6   \\
\hline
 \end{tabular}
 \caption{Relevant particle parameters relative to turbulent scales.}\label{table:particles}
\end{table}

Although Stokes-based particle relaxation time is not useful for predicting the dynamics of particles of this size,  we calculate it here as a point of reference.  The Stokes--flow relaxation time of the spherical particles is $\tau_{ps}^{(s)}= 3.64$ s.  The Stokes--flow relaxation time of the ellipsoidal particles is computed using the expression for the average drag coefficient of randomly oriented prolate ellipsoids found in Clift \emph{et al.} \cite{Clift:2005vm}:
\begin{equation}
C_{el}=3\pi d_e \frac{\sqrt{a^2-1}}{\log(a+\sqrt{a^2-1})},
\label{eq:d_el}
\end{equation}
where $a=l_e/d_e$ is the particle aspect ratio. Consequently, the particle response time for our ellipsoids of $a=2$ and $d_e=d_s$ becomes: 
\begin{equation}
\tau_{ps}^{(e)}=\tau_{ps}^{(s)}\: a \frac{\log(a+\sqrt{a^2-1})}{\sqrt{a^2-1}}=1.5\tau_{ps}^{(s)}
\label{eq:st_el}
\end{equation}
where $\tau_{ps}^{(s)}$ is based on the smaller (equatorial) diameter of the ellipsoid.  Our ellipsoidal particles thus have $\tau_{ps}^{(e)}=5.46$s.   According to \cite{Clift:2005vm}, we can define an equivalent diameter $d_{p}^{(eq)}$ for ellipsoidal particles as the diameter of a sphere having the same average drag coefficient as the ellipsoid. This can be determined from eq.~(\ref{eq:d_el}) to be $d_{p}^{(eq)}=10.5$ mm.   
Eq.~(\ref{eq:st_el}) offers an interesting insight on the effect of particle shape: particle response time is the ratio between inertial  and drag force; while this ratio scales with particle size for spheres, this is not true for elongated particle. For example, increasing particle size by  increasing the aspect ratio might actually decrease the response time and the equivalent diameter.

\subsection{Measurement techniques}
Fluid phase velocity measurements are performed using standard 2D particle image velocimetry (PIV).  These measurements use a single camera oriented perpendicular to a measurement volume of dimensions  4 cm x 3.4 cm x 0.1 cm centered in the tank. The final size of the interrogation area is 32$\times$32 pixels, corresponding to 0.8$\times$0.8 mm in physical space. This size is on the order of $2\eta$, thus spatial resolution effects are expected to be negligible \cite{Saarenrinne:2001gx}.  Statistics of the fluid phase are based on 510 independent velocity fields (corresponding to $>300$ integral time scales) containing 102$\times$86 datapoints each.  


We measure the kinematics of the large spherical and ellipsoidal particles by tracking the motion of tracers imbedded within them.  These tracers are made visible in a single planar cross--section of the particle whenever a particle moves (freely) through a laser light sheet in the $x-y$ plane.  To resolve the velocity in the $z$-direction, we employ stereoscopic PIV, and thus view the laser plane from two different directions, as seen in Figure \ref{fig:tank_setup}.  Using an algorithm described in \cite{Bellani:2012wn}, we use the 2D3C velocity fields inside the particle and the equation of solid body motion to determine both the linear and angular velocities of the particles (see Appendix for more details).  The optical setup is two cameras (Imager PRO-X, 1600 $\times$ 1200 pixels, both fitted with a 50 mm Nikkor lens and Scheimpflug/tilt adapter) viewing a 1mm thick laser light sheet,  capturing a measurement volume of  14 cm x 8 cm x 0.1 cm centered in the tank.  The two cameras view the measurement area from opposite sides, each at an angle of 35 degrees relative to the laser's forward-scatter direction. 
 Statistics of angular velocity are presented elsewhere \cite{Bellani:2012wn}, and are used herein primarily to control error in the linear velocity measurements (see Appendix). 

\section{Result}\label{sec:res}

 \begin{table}[tbp] 
\centering
\begin{tabular}{ccccccccccccc}                      
\hline
 &\emph{Input velocity:} &  $u_p^{(s)}$ & $u_p^{(e)}$ & $u_f$ & $\overline{(u_f)}_d$ &    $\overline{(u_f)}_l$ \\
 \hline
$\langle u^2 \rangle$ & $\times$10$^{-4}$ [m$^2$/s$^2$] &     1.29 &  1.06     &  1.55& 1.41   & 1.30  \\
& & [1.24\ 1.34]  & [1.01\ 1.10]  & [1.48\ 1.65]  & [1.35\ 1.48] & [1.25\ 1.36]\\
$\langle u^4 \rangle$/$\langle u^2 \rangle^2$ & [-] &     4.13           &      4.67     &  3.3  & 2.66  & 2.52  \\
& & [3.92\ 4.35]  & [4.41\ 4.95]  & [2.97\ 3.7]  & [2.51\ 2.83] & [2.41\ 2.62]\\
\hline
 \end{tabular}
 \caption{Second and fourth order statistics of particle and fluid phase velocity. The values in square brackets are the 95\% CI computed via the bootstrap method.}\label{table:vp_stats}
\end{table}

In this section we present statistics of particle velocity and compared them with fluid-phase velocity measurements. Particle velocity statistics are obtained  from 927 and 1076 Lagrangian trajectories for spheres and ellipsoids, respectively. 
For simplicity, we focus on the slip in the vertical direction, such that all data given for $u_f$ and $u_p$ are aligned with $y$.  
   
An important consideration to be made when comparing particle and fluid velocity statistics is whether or not particles show preferential concentration. When particles show preferential concentration, they selectively sample the fluid velocity field \cite{Eaton:2002uj,Gibert:2012dk,Fouxon:2012es}.  If they do, the particle statistics should be compared to fluid velocity statistics conditioned on vorticity \cite{Mortensen:2008bp}. Based on the results of \cite{Xu:2008dl}, we do not expect that our particles will behave in such a way, and thus we assume that they uniformly sample the velocity field. 

Although our measurement technique can provide simultaneous particle and fluid velocity measurements, we choose to compare particle velocity data to single phase measurements. This choice allows us to easily relate our results to other studies where statistics of inertial-particle motion are compared to the motion of tracer particles (\emph{e.g.} \cite{Qureshi:2007gx,Volk:2008fv}). It also allows a simple comparison with results from one-way coupled simulations such as those described in \cite{Calzavarini:2012fq}.

\begin{figure}[tbp]
\centering
\includegraphics[width=0.45\textwidth]{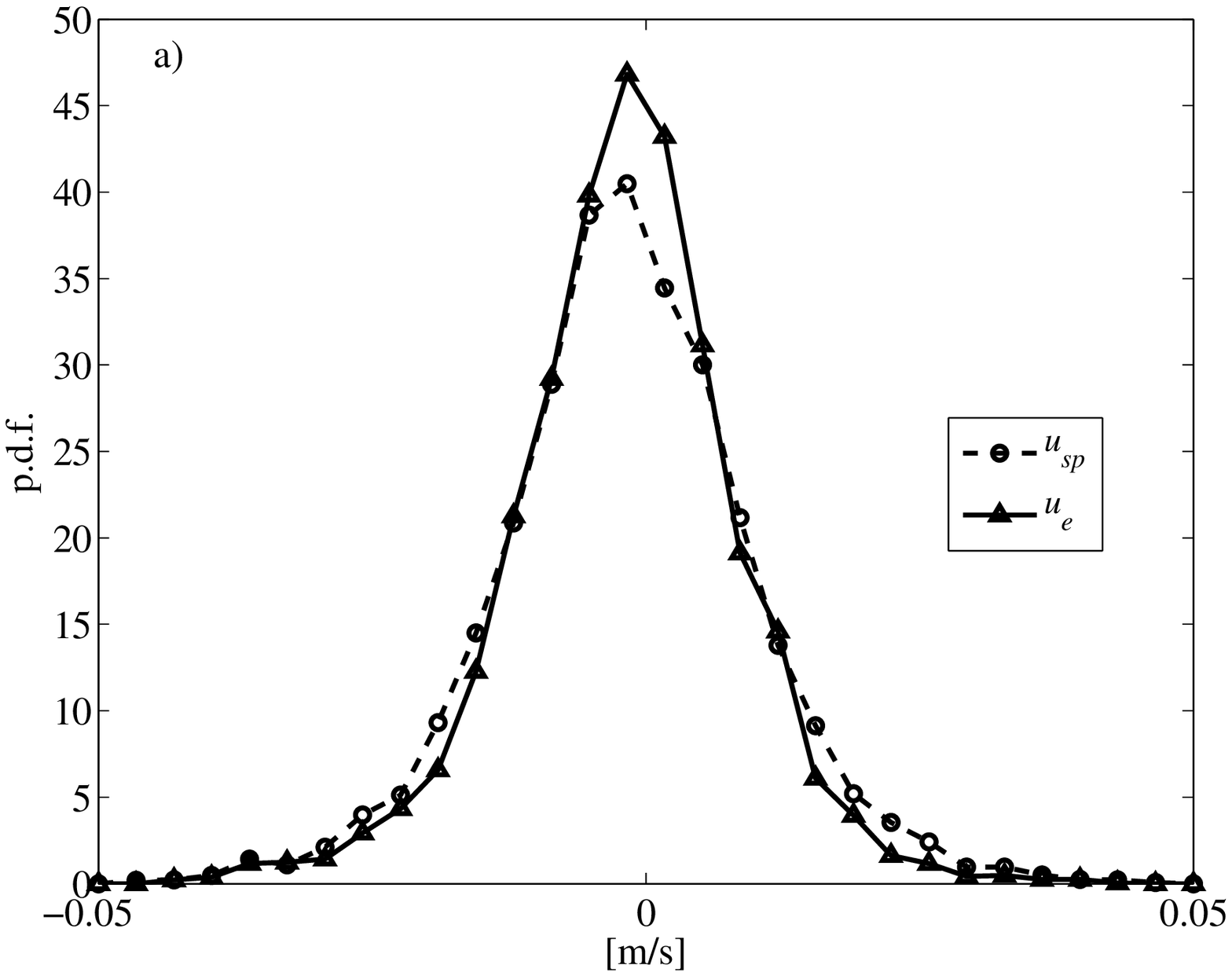}
\includegraphics[width=0.45\textwidth]{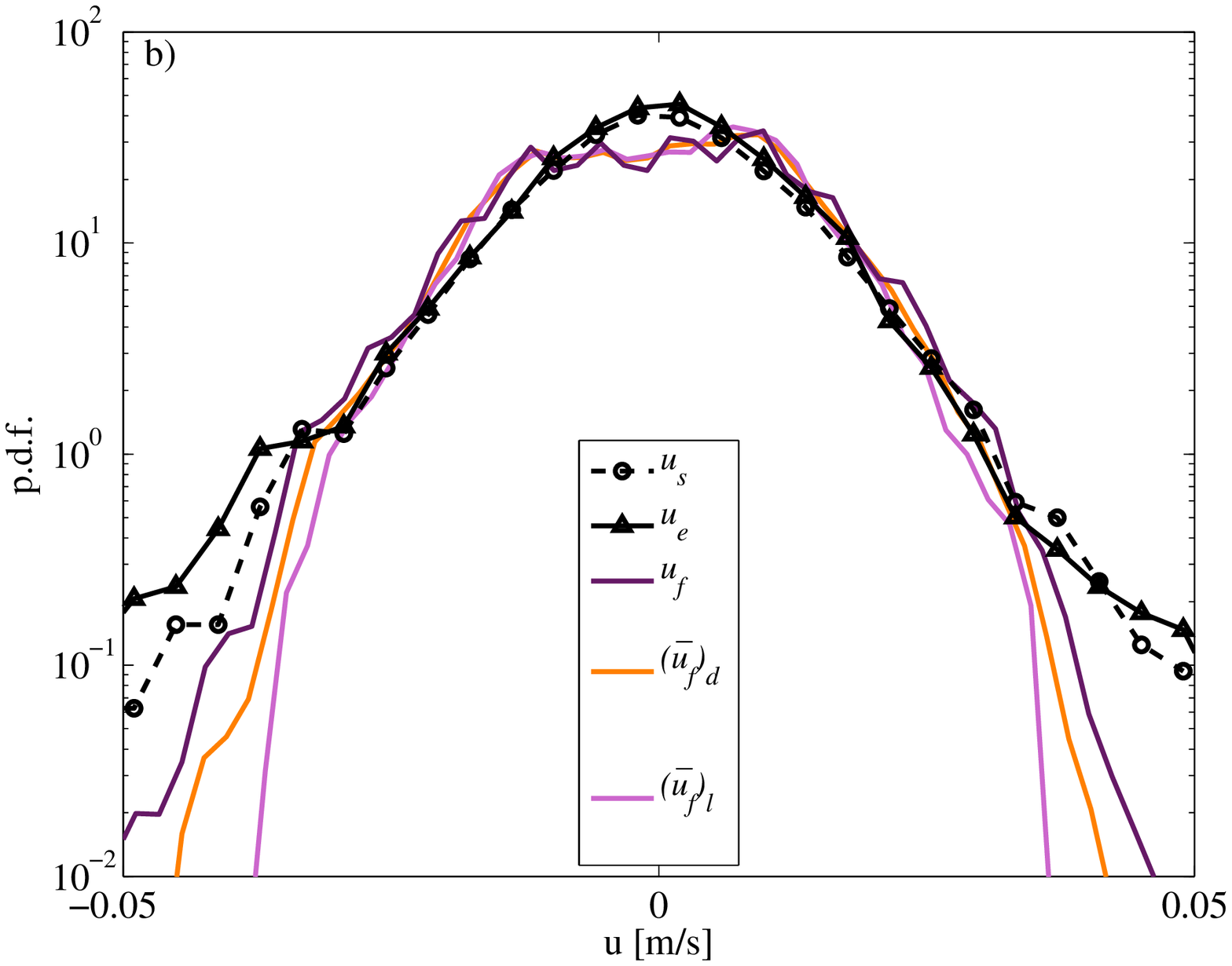}\\
\includegraphics[width=0.45\textwidth]{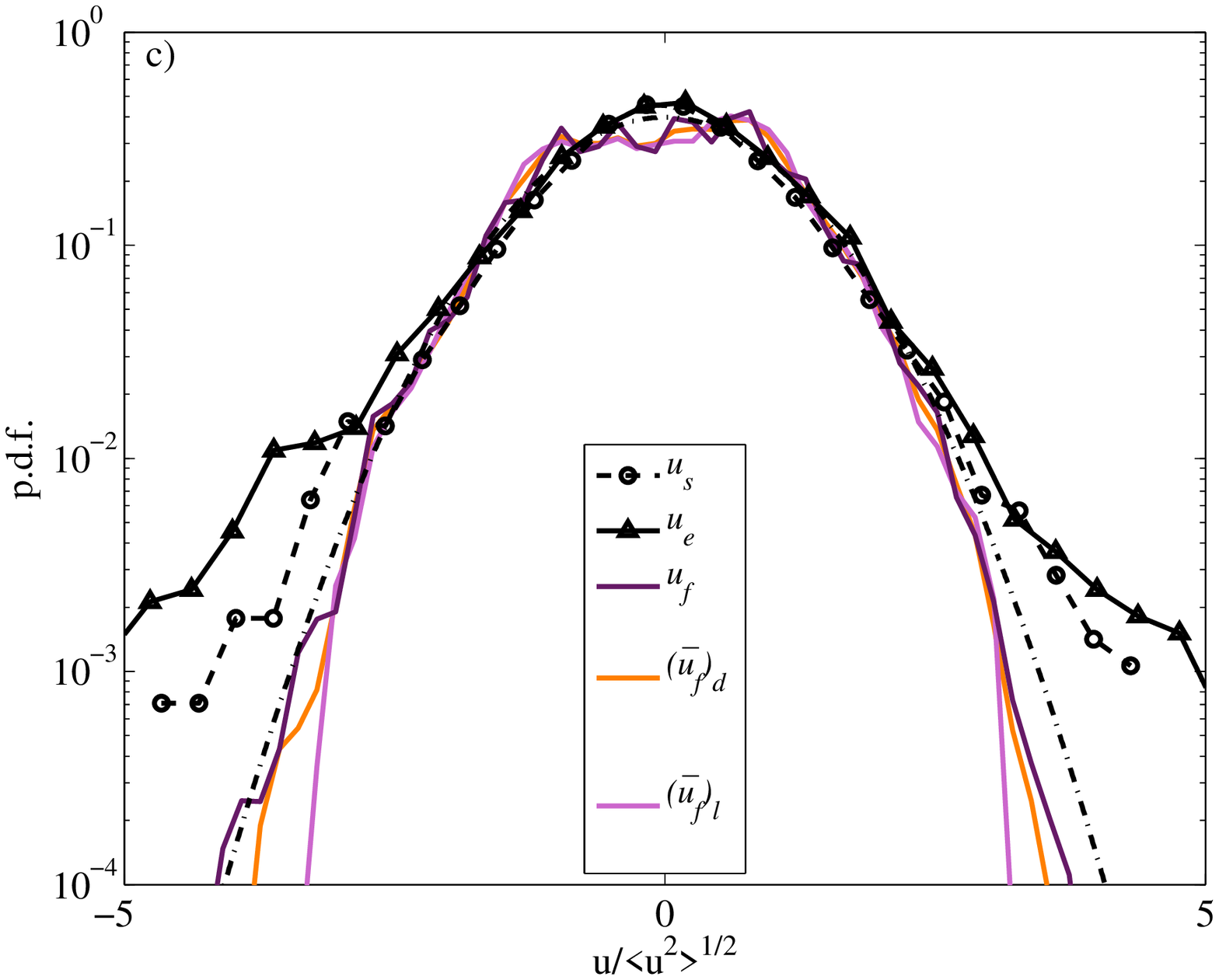}
\caption{a) Histograms of spherical (solid lines) and ellipsoidal (dashed lines) particles' velocity components. b) Comparison with fluid velocity \emph{pdf}s. c) Standardized pdf showing hyper-gaussianity of particle velocity (dot--dash line is Gaussian). The figure shows that particle velocity is not trivially related to velocity  at any scale.}\label{fig:Hist}  
\end{figure}

 The probability density functions (\emph{pdf}s) of linear velocities ($u_p$) are shown in figure \ref{fig:Hist}a.  For both spheres and ellipsoids, the \emph{pdf}s are symmetric around $u_p=0$, as expected for homogeneous isotropic turbulence with zero mean flow.  Table \ref{table:vp_stats} shows that both particle types have velocity variance that is significantly smaller than that of the fluid phase.  This is an immediate indication that particles do not behave as passive tracers.  The variance of ellipsoidal particle velocities is significantly lower than that for spheres, implying that they follow the flow less faithfully than the spheres, likely due to their shape or to their longer relaxation timescale (ellipsoids' Stokes--based response time is 50\% longer than spheres, as derived in section \ref{sec:exp}).    

Figures \ref{fig:Hist}b and \ref{fig:Hist}c compare the particle velocity \emph{pdf}s to the fluid phase velocity.  We also include \emph{pdf}s of the velocity field filtered at lengthscales corresponding to the spherical particle diameter ($\overline{(u_f)}_d$) and the ellipsoids' major axis ($\overline{(u_f)}_l$). Comparison with these demonstrates the nonlinear dynamics of particle-turbulent interactions, as discussed below. 
  
The shape of the standardized \emph{pdf}s seen in figure \ref{fig:Hist}c is of particular interest, because it reveals whether particles simply act as a linear filter of the velocity field, or if the interaction is more complex.  From figure \ref{fig:Hist}c we can see that while fluid velocity follows a Gaussian or sub-Gaussian distribution, particles show a super-Gaussian behavior.  This behavior is especially strong for ellipsoidal particles. We quantify this effect with fourth order moments of the \emph{pdf}s, see table (\ref{table:vp_stats}). This indicates that the filtering relationship between particle and fluid velocity is nonlinear, as is discussed further in section \ref{sec:dis}.
   
Applying the definition of slip velocity given in eq.~(\ref{eq:usl2}) we obtain that slip velocity for spherical particles is $V^{(s)}=  5.5$ mm/s and for ellipsoidal particles is $V^{(e)}=7.1$ mm/s. 
The definition of stochastic slip velocity $V$ can also be modified to so that it uses the variance of a spatially filtered fluid velocity field.  The advantage of doing this is that it could explicitly represent the manner in which particles spatially filter the fluid velocity field.  Doing so also provides an unambiguous definition for the covariance term in $V_B$.  However, it is not yet known what filter best approximates the particles' experience of the fluid velocity field.  Indeed, if this were known, it would be an important step forward in modeling fluctuating drag on large particles in turbulence.  Here, we use a simple boxcar filter, and find that, although the slip velocity decreases, as expected, with filter length, its value never vanishes, even for filter length of the size of the particles scale or larger. This is another indication that particle motion is dominated by scales much larger than the particle's size.  

We can use the stochastic slip velocity $V$ to define a Reynolds number $Re_V\equiv Vd_p\nu^{-1}$.  For the particles measured here, $Re_V$ is 46 for spheres and 79 for ellipsoids.  
We consider this result the first indication that $Re_V$ may indicate the degree of nonlinearity in the particle--fluid interactions.  This hypothesis is driven by two observations.  First, as $Re_V$ approaches zero, the fluid particles follow the flow as passive tracers, and thus their motion is a linear function of the fluid phase velocity.  Second, the particles measured here have $Re_V>1$, demonstrate a nonlinear response to the fluid velocity field (as indicated by the \emph{pdfs}), and the strength of this nonlinearity is correlated with the Reynolds number (\emph{i.e.} ellipsoids have a larger $Re_V$ and exhibit greater nonlinearity).

 \begin{table}[tbp] 
\centering
\begin{tabular}{ccccc}                      
\hline
  & \emph{Slip velocity}  $V$ &: $(\langle u_f^2\rangle-\langle u_p^2\rangle)^{1/2}$         &     $(\langle \overline{(u_f)}_d^{2} \rangle-\langle u_p^2\rangle)^{1/2}$    &  $(\langle \overline{(u_f)}_l^{2} \rangle-\langle u_p^2\rangle)^{1/2}$ \\
 \hline
\emph{Spheres} & $\times$10$^{-3}$ [m/s] &     5.5    (Re$_p$=46)       &      3.5  (Re$_p$=29)   &  1 (Re$_p$=8)   \\
\emph{Ellipsoids} & $\times$10$^{-3}$ [m/s] &     7.1   (Re$_p$=79)        &      5.9   (Re$_p$=66)  &  4.9 (Re$_p$=46) \\
\hline
 \end{tabular}
 \caption{Slip velocities and particle Reynolds numbers.}\label{table:vp_slip}
\end{table}

\section{Discussion}\label{sec:dis}
The kurtosis and scaled \emph{pdf}s of particle velocities show that large particles like the ones used here exhibit super--Gaussian behavior.  Super--gaussianity of large spherical particles was also reported by  Qureshi \emph{et al.} \cite{Qureshi:2007gx}, specifically in the acceleration statistics. In their experiments, intermittent behavior persists for particle sizes up to the largest turbulent scales.  As in our experiments, they found that the fluid statistics remained nearly Gaussian, in marked contrast to the particle statistics.

The evidence of super--Gaussian behavior is important, in that it is a signature of intermittency.  Intermittency is a well--known feature of turbulent processes (often at small scales and in acceleration statistics) \cite{Tsinober:2001vv,BIFERALE:2004hj,Bec:2006eq}, and can provide a clue as to the dynamics of particle--fluid interaction.  For example, if particles simply followed a linear filter of the fluid phase velocity, then we would not expect intermittent behavior.  Based on their results, Qureshi \emph{et al.} suggest that particle acceleration statistics may be related to statistics of the Eulerian pressure increment.  Such a coupling could improve the final term in eq.~(\ref{eq:MR}), though the work of \cite{Bagchi:2003ey} and \cite{Calzavarini:2012fq} emphasize that the drag term requires improvement as well.

Calzavarini \emph{et al.} \cite{Calzavarini:2012fq} use a nondimensional version of our proposed slip velocity ($V^\ast\equiv V_A^2\langle u_f\prime^{-2}\rangle$) to evaluate how well their Euler--Lagrange model reproduces particle velocity statistics.  They compare with DNS data for particles as large as $d_p=15\eta$.  We can add our result for spheres of $d_p=21\eta$ to this analysis, helping evaluate drag models. A useful observation is made by Sapsis \emph{et al.} \cite{Sapsis:2011ii}, who found that $V_B^2\langle u_f\prime^{-2}\rangle$ is essentially constant with turbulence Reynolds number, making it easy to compare $V^\ast$ across different experiments. Our data ($V^\ast= 0.16$ [0.12 0.20]) and those in \cite{Sapsis:2011ii},  compared with the results discussed in \cite{Calzavarini:2012fq} and \cite{HOMANN:2010ey}, support the possibility  of non-monotonic behavior of $V^\ast$ with $d_p$.

A unique aspect of our measurements is that we are able to compare the slip velocity for particles of different shape.  
The drag models discussed above are valid for spherical particles. For ellipsoidal particles, non-linear extensions of creeping flow solutions are not available. Such extensions are complicated by the fact that non--spherical particles have orientation--dependent drag.  As a result, ellipsoidal particles can have significantly different wake structures than spheres of similar size.  This was observed in numerical simulations \cite{ELKHOURY:2010hb}, in which ellipsoids produced more energetic wakes than spheres.  We see evidence of this type of shape--dependent wake energy in our data, namely in that ellipsoids show a stronger deviation from Gaussian velocity statistics than spheres do.  This enhanced intermittency may depend on size as well as shape, given that the ellipsoids are slightly larger than the spheres.

At this stage, we see three main uses for our definition of stochastic slip velocity $V\equiv {\left( {\langle {u_f'^2}\rangle  - \langle {u_p'^2}\rangle } \right)^{1/2}}$.  The first is to quantify the importance of the drag due to the fluctuating relative velocity.  When $V=0$, the particle acts as a tracer and thus the fluctuating component is zero. In this regime standard empirical coefficients are expected to work well, and an instantaneous slip velocity can be determined unambiguously.   When $V$ is nonzero, we can compare it to the steady slip velocity, $V_s$, to understand the relative importance of steady and fluctuating slip.  When $V$ is large compared to its steady counterpart, we can expect that empirical values used in the standard drag model will not be applicable. 

Thus the second major use for $V$ is as an input for formulating stochastic drag models for the fluctuating part of the drag for arbitrarily shaped particles. Such drag models could include empirical coefficients based on the stochastic particle Reynolds number $Re_V\equiv {\left( {\langle {u_f'^2}\rangle  - \langle {u_p'^2}\rangle } \right)^{1/2}}d_p\nu^{-1}$ and the steady particle Reynolds number $Re_S\equiv {\left( {\langle {u_f}\rangle  - \langle {u_p}\rangle } \right)}d_p\nu^{-1}$.  While $Re_S$ can be inserted into the standard drag model \cite{Clift:2005vm}, more data is needed before we can easily interpret the flow behavior implied by $Re_V$, and this progress is closely tied to the development of stochastic drag models employing $V$.

The third major use for $V$ is in predicting some of the key features of particle-laden turbulent suspensions.  There is currently no method for predicting turbulence modulation (such as TKE amplification or attenuation) and particle clustering (\emph{i.e.} preferential concentration) for suspensions of large particles in turbulence.  For small particles, the Stokes number has been used to predict these effects, but it fails for large particles as discussed above \cite{Lucci:2011dx}.  We expect that whatever non-dimensional parameters predict clustering and TKE modulation for large particles will also control the stochastic slip velocity $V$.  If this is true, then $V$ could serve as a midway predictor for clustering and TKE modulation.  This would serve two purposes: $V$ may be easier to measure
than the clustering and TKE modulation, and the relationship between $V$ and stochastic drag, once found, would guide the search for a predictor of clustering and TKE modulation.

\section{Conclusions}
The standard drag model fails to accurately describe forces on large particles in turbulent flows.  We can replace the standard drag with the sum of steady and fluctuating drag terms, but the fluctuating drag term has not yet been parameterized.  Herein, we discuss the definition of a stochastic slip velocity.  This can serve as an input to fluctuating drag models, and also serve other key analyses in turbulent particle suspension dynamics.  Specifically, it allows us to determine a particle Reynolds number for neutrally buoyant particles, and contributes to the extension of the Stokes number.  This is essential for efforts to accurately predict the dynamics of large particles in turbulent flows.

We discuss two options for the stochastic slip velocity, and demonstrate their approximate equivalence.  We then measure the stochastic slip velocity using laboratory data collected on large ellipsoidal and spherical particles in homogeneous isotropic turbulence.  The results suggest that particle Reynolds number based on the stochastic slip is an effective predictor of nonlinear fluid--particle interactions.  Specifically, both particles show velocity distributions with greater kurtosis than the fluid phase velocity distribution.  Comparing the ellipsoids and spheres, we see that the increase in kurtosis is correlated with an increase in particle Reynolds number.

A key next step is to obtain the necessary measurements with which to quantify the covariance between velocity and fluid fluctuations, and thus evaluate the difference between our two proposed definitions of stochastic slip velocity ($V_A$ and $V_B$).  In addition, we can evaluate rotational slip by using within--particle velocities to extract particle rotation measurements.

\section*{Appendix}
We use stereoscopic PIV to measure the velocities $\mathbf{U}_M$ and $\mathbf{U}_N$ at two points within the particle: $\mathbf{X}_M$ and $\mathbf{X}_N$. PIV measurements are computed by the commercial software \emph{Davis} from \emph{Lavision Gmbh}, whose accuracy is discussed in detail in \cite{Stanislas:2008ht}. The details of the PIV settings used in tese measurements are discussed in \cite{Bellani:2012wn}. From these measurements the angular velocity $\mathbf{\Omega}$ can be determined from the equation of solid body rotation:
\begin{equation}
\mathbf{U}_N=\mathbf{U}_M+\Omega \times (\mathbf{X}_N-\mathbf{X}_M).
\label{eq:rot_method}
\end{equation}
Since PIV measurements are in a single $x-y$ plane, eq.~(\ref{eq:rot_method}) is over-determined in $\Omega_z$ and under-determined in $\Omega_x$ and $\Omega_y$.  By including a third point $P$ in addition to $M$ and $N$, all three components of $\mathbf{\Omega}$ can be determined. 

Because PIV provides many more than 3 velocities within each particle, we can improve the precision and accuracy of $\mathbf{\Omega}$ measurements.  We do this by calculating a value of $\mathbf{\Omega}$ using each possible set of three points $M$, $N$ and $P$, and considering the distribution of $\mathbf{\Omega}$ estimates.

The location of the particle center of gravity is not explicitly measured by PIV, but the velocity there ($\mathbf{V_{cg}}$) can be determined as follows.  The equation of solid body rotation gives:
\begin{equation}\label{eq:rot1}
\mathbf{V_{cg}}=\mathbf{V_{cf}}-\mathbf{\Omega}\times \mathbf{r},
\end{equation}
where  $\mathbf{V_{cf}}$ is the velocity at the center of the particle cross-section formed by the intersection of the particle and the PIV measurement plane, $\mathbf{\Omega}$ is the angular velocity as calculated above, and $\mathbf{r}=(r_x,r_y,r_z)$ is the distance between the center of gravity of the particle and the center of the particle cross-section. Note that for spherical particles $r_x=r_y=0$.
From eq.~(\ref{eq:rot1}), we can express the variance of $\mathbf{V_{cg}}$ as:  
\begin{equation}\label{eq:rot2}
\langle\mathbf{V_{cg}}^2\rangle=\langle\mathbf{V_{cf}}^2\rangle-\langle(\mathbf{\Omega}\times \mathbf{r})^2\rangle-\langle2\mathbf{V_{cg}}(\mathbf{\Omega}\times \mathbf{r})\rangle.
\end{equation}
It is reasonable to assume that the last term on the RHS of eq.~\ref{eq:rot2} vanishes because the velocity of the center of gravity and the velocity induced by particle rotation at the measurement location are uncorrelated. 
We therefore are left with the expression:
\begin{equation}\label{eq:rot3}
\langle\mathbf{V_{cg}}^2\rangle=\langle\mathbf{V_{cf}}^2\rangle-\langle(\mathbf{\Omega}\times \mathbf{r})^2\rangle.
\label{eq:Vcg}
\end{equation}
This indicates that the variance of the velocity at the particle's center of gravity is the variance of the velocity measured at the center of a randomly oriented plane intersecting the particle minus the `rotation--induced noise� term  $\langle(\mathbf{\Omega}\times \mathbf{r})^2\rangle$ that is due to the velocity induced by particle rotation at the measurement point.

To quantify the importance of the rotation--induced noise term, we expand  $\mathbf{V_{rin}}=\mathbf{\Omega}\times \mathbf{r}$ in its components.  Because spherical particles have $r_x=r_y=0$, the noise term becomes $(u_{rin},v_{rin},w_{rin})=(\Omega_y r_z,-\Omega_x r_z, 0)$. Therefore, we can immediately conclude that the variance of $w_{cf}$ equals the variance of $w_{cg}$.  The variance of the other two components can be expanded by noting that the vector $\mathbf{\Omega}$ is independent of $r_z$.  Therefore we can express the variance of the products as the product of the variances: $\langle(\Omega_i r_z)^2\rangle=\langle\Omega_i^2\rangle\langle r_z^2\rangle$.  
The variance of the distance $r_z$ of a point chosen randomly within the volume of a spherical particle of radius $R$ is $\langle r_z^2\rangle = \frac{3}{5}R^2$. Therefore $\langle u_{rin}^2 \rangle=\frac{3}{5}R^2\langle \Omega_x^2 \rangle$, and  $\langle v_{rin}^2 \rangle=\frac{3}{5}R^2\langle \Omega_y^2 \rangle$. 

In our measurements, the diameter of spherical particles is $R=4\times10^{-3}$ m, and the angular velocity measurements show that $\langle\Omega_i^2\rangle=O(10^{-1})$. Therefore the noise term is $O(10^{-7})$.  Comparing this to $\langle u_{cf}^2 \rangle$, which is $O(10^{-4})$, we conclude that the rotation--induced noise is negligible in our measurements of spherical particles� velocity variance.  For ellipsoidal particles, the expression for 
$\mathbf{V_{rin}}$ becomes much more complicated because $r_x$ and $r_y$ are non-zero and the expression for the radius variance is different than that for spheres.  However, our measurements of $\mathbf{\Omega}$ for ellipsoids are very similar to the results for spheres.  Given this, and the fact that the ellipsoids are very similar size to the spheres, we expect the order of magnitude of the noise term to be similar for spheres and ellipsoids.  As a result, we neglect the rotation--induced noise when calculating the variance of ellipsoid velocities.  Should these terms be non--negligible in other cases, they can be quantified using the method presented here.  
To further limit the impact of measurement noise on fluid phase and particle statistics, we pre-process the PIV data using the median test proposed by \cite{Westerweel:2005dy}, and post-process the time-series of particle velocity obtained from eq.~(\ref{eq:Vcg}) with a standard median test.  

\section*{References}
\bibliographystyle{unsrt}
\bibliography{new_cit}
\end{document}